\documentclass{aa}  

\newcommand{\Msun}{\ensuremath{M_\odot}}
\newcommand{\Rsun}{\ensuremath{R_\odot}}
\newcommand{\Zsun}{\ensuremath{Z_\odot}}
\newcommand{\Lsun}{\ensuremath{L_\odot}}
\newcommand{\days}{\ensuremath{\mathrm{days}}}
\newcommand{\yr}{\ensuremath{\mathrm{yr}}}
\newcommand{\kpc}{\ensuremath{\mathrm{kpc}}}
\newcommand{\kelvin}{\ensuremath{\mathrm{K}}}
\newcommand{\POSYDON}{\texttt{POSYDON}\xspace}
\newcommand{\MESA}{\texttt{MESA}\xspace}
\newcommand{\Gaia}{\texttt{Gaia}\xspace}

\usepackage{graphicx}
\usepackage{txfonts}
\usepackage{natbib}
\usepackage[colorlinks=true,
    linkcolor=blue,
    filecolor=magenta,      
    urlcolor=cyan,
    citecolor=blue]{hyperref}
\usepackage{scrextend}

\begin{document} 

 \title{The Formation of Black Holes in Non-interacting, Isolated Binaries}
 \titlerunning{Black Holes in Non-interacting Binaries}
 \subtitle{\Gaia Black Holes as Calibrators of Stellar Winds From Massive Stars}

 \author{Matthias~U.~Kruckow \inst{\ref{inst:Geneva}, \ref{inst:GWSC}}\thanks{Matthias.Kruckow@unige.ch}
         \and
         Jeff~J.~Andrews \inst{\ref{inst:Florida}, \ref{inst:Florida2}}
         \and
         Tassos~Fragos \inst{\ref{inst:Geneva}, \ref{inst:GWSC}}
         \and
         Berry~Holl \inst{\ref{inst:Geneva}, \ref{inst:Ecogia}}
         \and\\
         Simone~S.~Bavera \inst{\ref{inst:Geneva}, \ref{inst:GWSC}}
         \and
         Max~Briel \inst{\ref{inst:Geneva}, \ref{inst:GWSC}}
         \and
         Seth~Gossage \inst{\ref{inst:CIERA}}
         \and
         Konstantinos~Kovlakas \inst{\ref{inst:ICE}, \ref{inst:IEEC}}
         \and
         Kyle~A.~Rocha \inst{\ref{inst:Northwestern}, \ref{inst:CIERA}}
         \and
         Meng~Sun \inst{\ref{inst:CIERA}}
         \and
         Philipp~M.~Srivastava \inst{\ref{inst:CIERA}, \ref{inst:Comp}}
         \and
         Zepei~Xing \inst{\ref{inst:Geneva}, \ref{inst:GWSC}, \ref{inst:CIERA}}
         \and
         Emmanouil~Zapartas \inst{\ref{inst:FORTH}, \ref{inst:IAASARS}}
        }

 \institute{\label{inst:Geneva}D\'epartement d'Astronomie, Universit\'e de Gen\`eve, Chemin Pegasi 51, CH-1290 Versoix, Switzerland
            \and
            \label{inst:GWSC}Gravitational Wave Science Center (GWSC), Universit\'e de Gen\`eve, 24 quai E. Ansermet, CH-1211 Geneva, Switzerland
            \and
            \label{inst:Florida}Department of Physics, University of Florida, 2001 Museum Rd, Gainesville, FL 32611, USA
            \and
            \label{inst:Florida2}Institute for Fundamental Theory, 2001 Museum Rd, Gainesville, FL 32611, USA
            \and
            \label{inst:Ecogia}D\'epartement d'Astronomie, Universit\'e de Gen\`eve, Chemin d'Ecogia 16, CH-1290 Versoix, Switzerland
            \and
            \label{inst:CIERA}Center for Interdisciplinary Exploration and Research in Astrophysics (CIERA), 1800 Sherman, Evanston, IL 60201, USA
            \and
            \label{inst:ICE}Institute of Space Sciences (ICE, CSIC), Campus UAB, Carrer de Magrans, 08193 Barcelona, Spain
            \and
            \label{inst:IEEC}Institut d’Estudis Espacials de Catalunya (IEEC), Edifici RDIT, Campus UPC, 08860 Castelldefels (Barcelona), Spain
            \and
            \label{inst:Northwestern}Department of Physics \& Astronomy, Northwestern University, 2145 Sheridan Road, Evanston, IL 60208, USA
            \and
            \label{inst:Comp}Electrical and Computer Engineering, Northwestern University, 2145 Sheridan Road, Evanston, IL 60208, USA
            \and
            \label{inst:FORTH}Institute of Astrophysics, FORTH, N. Plastira 100, Heraklion, 70013, Greece
            \and
            \label{inst:IAASARS}IAASARS, National Observatory of Athens, Vas. Pavlou and I. Metaxa, Penteli, 15236, Greece
           }
 \authorrunning{M.~U.~Kruckow et al.}

 \date{Received Month Day, Year; accepted Month Day, Year}

 \abstract
  {The black holes discovered using \Gaia, especially \Gaia BH1 and BH2, have low mass companions of solar-like metallicity in wide orbits. For standard isolated binary evolution formation channels including interactions such an extreme mass ratio is unexpected; especially in orbits of hundreds to thousands of days.}
  {Here, we investigate a non-interacting formation path for isolated binaries to explain the formation of \Gaia BH1 and BH2.} 
  {We use single star models computed with \MESA to constrain the main characteristics of possible progenitors of long-period black hole binaries like \Gaia BH1 and BH2. Then, we incorporate these model grids into the binary population synthesis code \POSYDON, to explore whether the formation of the observed binaries at solar metallicity is indeed possible.}
  {We find that winds of massive stars ($\gtrsim 80\,\Msun$), especially during the Wolf-Rayet phase, tend to cause a plateau in the initial stellar mass to final black hole mass relation (at about $13\,\Msun$ in our default wind prescription). However, stellar winds at earlier evolutionary phases are also important at high metallicity, as they prevent the most massive stars from expanding ($<100\,\Rsun$) and filling their Roche lobe. Consequently, the strength of the applied winds affects the range of the final black hole masses in non-interacting binaries, making it possible to form systems similar to \Gaia BH1 and BH2.}
  {We deduce that wide binaries with a black hole and a low mass companion can form at high metallicity without binary interactions. There could be hundreds of such systems in the Milky Way. The mass of the black hole in binaries evolved through the non-interacting channel can potentially provide insights into the wind strength during the progenitors evolution.}


\maketitle

\section{Introduction}
\label{sec:Intro}

Traditionally, it is difficult to observe non-accreting black holes (BHs) because of their non emissivity in the electromagnetic spectrum. The first indirect measurements of stellar mass BHs are from X-ray binaries, where hot material is accreted onto an unseen but very compact object \citep[e.g.][]{1972Natur.235...37W}. Those binaries are well observed due to their high-energy emission \citep[e.g.][and references therein]{2006ARA&A..44...49R}. The compact object itself is usually hidden behind the strong emission of the accreting material, which can outshine the donating companion, too. Additionally, the best studied stellar X-ray sources are traditionally found to be Galactic \citep{2014SSRv..183..223C}.

In recent years, observations of gravitational waves showed a broader existence of BHs in binaries throughout the Universe. Those observations question our understanding of BHs because of their large variety of BH masses \citep{GWTC-1, GWTC-2, GWTC-3, LVK-pop, GWTC-2.1}, while we infer only moderate BH masses for X-ray sources \citep{2010ApJ...725.1918O, 2011ApJ...741..103F}.

Most recently, \Gaia DR3 data has been searched for unseen massive companions to stars in the Milky Way (MW). These investigations revealed some candidates, named \Gaia BH1 \citep{chakrabarti2023, el-badry-BH1}, \Gaia BH2 \citep{el-badry-BH2, tanikawa-BH2}, \Gaia BH3 \citep[][using pre-release DR4 data]{Gaia-BH3}, as well as other possible candidates of unseen compact objects \citep[e.g.][]{2022arXiv220700680A, 2024arXiv240906352S}. Follow-up radial velocity measurements have verified three individual \Gaia BHs. \Gaia BH3 \citep{Gaia-BH3} is rather peculiar among BHs in the MW in term of its mass, while within expectations from gravitational waves. It has a mass of $32.7\pm0.82\,\Msun$, and is believed to be formed within a tidally disrupted globular cluster at lower metallicity \citep{balbinot2024, el-badry2024, iorio2024, marin2024}. In contrast, \Gaia BH1 and BH2 have masses of $9.62\pm0.18\,\Msun$ and $8.94\pm0.34\,\Msun$, respectively, and are both in the MW disk with roughly solar metallicity companion stars. Some other formation scenario is required.

From classical binary evolution, one expects binaries consisting of a BH with a stellar companion in orbits up to a few thousands of days \citep{2017ApJ...850L..13B, 2018MNRAS.481..930Y, 2022ApJ...931..107C}. The binaries in those studies favour companion masses above $1\,\Msun$ \citep{2017MNRAS.470.2611M, 2018ApJ...861...21Y}. 
Depending on their mass, OB or Wolf-Rayet star companions can accrete mass stably upon Roche lobe overflow from the BH's progenitor \citep{2019A&A...627A.151S, 2020A&A...638A..39L}. Even if this mass transfer becomes unstable, these binaries are more likely to survive. Low mass companions on the other hand, like those in \Gaia BH1 and BH2, will most certainly experience unstable mass trasnfer and either merge within a common envelope, because they lack the orbital energy to successfully eject a common envelope, or result in tight orbits, $<100\,\days$, like the progenitors of low mass X-ray binaries \citep[e.g.][]{2003MNRAS.341..385P,2015ApJ...800...17F}. Previous studies have suggested \Gaia BH1 and BH2 were formed dynamically in an open cluster \citep{rastello2023, di_carlo2024, tanikawa2024}. However, this is not the only possible formation scenario. In their discovery paper for \Gaia BH1, \citet{el-badry-BH1} propose four other possible scenarios including classical binary evolution involving at least one mass transfer phase, triple system dynamics \citep[e.g.][]{2024ApJ...964...83G}, and formation in a hierarchical triple \citep[e.g.][]{2024arXiv241010581L}. Here, we will explore in depth, one of those possible scenarios: the idea that the progenitors of \Gaia BH1 and BH2 were initially very massive stars ($M>50\,\Msun$) as recently studied by \citet{2024MNRAS.535L..44G}. Such massive stars have so extreme wind mass loss rates on the main sequence that they lose their entire envelopes prior to core hydrogen exhaustion, thus never expanding onto the giant branch. Hence, these systems avoid mass transfer. Subsequent wind mass loss during their evolution as Wolf-Rayet stars further reduces their initially high masses \citep[see, e.g.,][]{bavera2023, 2024arXiv240707204V}. The resulting BHs are of sufficiently low mass that they could potentially explain the BHs in \Gaia BH1 and BH2.

Our approach is to use the \POSYDON binary population synthesis framework \citep{2023ApJS..264...45F, 2024ApJ...........A} to evolve a population of stellar binaries with extreme mass ratios. The primary star is sufficiently massive that it never expands into a giant star, while the companion is $\lesssim 1\,\Msun$ to explain the luminous companions to \Gaia BH1 and BH2. In Section~\ref{sec:Methods}, we describe our adaptations to \POSYDON and associated single-star models calculated with \MESA, made specifically to model \Gaia BHs. In Section~\ref{sec:Results}, we provide our results from our single-star model grids and binary population-synthesis models. Finally, in Section~\ref{sec:Discussion}, we discuss our results and provide some conclusions.

\section{Methodology}
\label{sec:Methods}

All results presented in this paper are derived using the binary population synthesis framework \POSYDON \citep{2023ApJS..264...45F, 2024ApJ...........A}. \POSYDON uses extensive, detailed single- and binary-star model grids, calculated with the \MESA stellar structure and binary evolution code \citep{2011ApJS..192....3P, 2013ApJS..208....4P, 2015ApJS..220...15P, 2018ApJS..234...34P, 2019ApJS..243...10P} to self-consistently follow the entire evolution of both stars in a stellar binary. Specifically for this work, we employ the \POSYDON infrastructure to calculate additional single-star grids with \MESA. We analyse them with \POSYDON tools and use \POSYDON to perform population-synthesis calculations.

\subsection{Single Star Models}
\label{sec:single}
In addition to the default single-stars grids in \POSYDON{}~v2 \citep{2024ApJ...........A}, covering different metallicities, we generate several extra models at solar metallicity \citep[we use $\Zsun=0.0142$ following][]{2009ARA&A..47..481A} with variations on our default wind prescriptions, presented in Section~\ref{sec:Wind}. \POSYDON modifies the default wind scheme in \MESA and will be described in Appendix~\ref{sec:wind_prescriptions}.

While multiple wind prescriptions can affect stars throughout their lifetimes depending on their evolutionary state, we find that for massive stars there are two dominating winds, which we call Vink and WR winds throughout the paper. The Vink wind \citep{2001A&A...369..574V} dominates the early evolution of the massive stars during the core-H burning, while the WR wind \citep{2000A&A...360..227N} is usually optically thick, and kicks in after the star loses most of its H-rich envelope. Therefore, we test models in which both of these winds are altered with multiplicative factors. Separately, winds are well-known to be affected by metallicity. Therefore, although we focus on models with $\Zsun$ (observations suggest the companions to \Gaia BH1 and BH2 have $\lesssim\Zsun$), we additionally show models with $0.45\,\Zsun$ and $2\,\Zsun$ without additional modifications in Section~\ref{sec:Wind}.

\subsection{Binary Population Synthesis Models}
\label{sec:BPS}
To compare our single star models to an observed population we use \POSYDON to simulate the intrinsic population of non-interacting binaries with a low-mass star ($0.5\text{ to }2.0\Msun$) orbiting a BH. We modify the default setup of \POSYDON by adding a user module\footnote{see \url{https://github.com/mkruckow/POSYDON/tree/matthias_Gaia_BH}} to allow for the modelling of extreme, initial mass ratio binaries (mass ratio $<0.05$). Specifically, we modify the ``flow chart'' to start our simulations in the ``detached step'' \citep[solving a simplified set of ordinary differential equations to model the effects of wind mass loss, tides, magnetic braking, and gravitational radiation on the orbital and rotation parameters; for more details see Section~8.1 in][]{2023ApJS..264...45F} to make use of our various single star grids and allow initial, non-zero eccentricity at zero-age main sequence (ZAMS).

Our population-synthesis models contain 1 million binaries each. These binaries are generated in a starburst and weighted in a post-processing step to resemble the MW, assuming a constant star formation history. For the results presented in Section~\ref{sec:Pop-syn} we use a thermal eccentricity distribution at ZAMS\footnote{\label{fn:default}In Appendix~\ref{sec:Pop-syn2} we use the \POSYDON defaults: circular orbits at ZAMS, BH formation following \citep{2020MNRAS.499.2803P}, and supernova kicks scaled by one over the BH mass.}. To focus on potential progenitors of \Gaia BH1 and BH2-like systems, we only sample primary masses $\in [20,150]\,\Msun$ and secondary masses $\in [0.5,2.0]\,\Msun$. Since we lack robust observational constraints for these extreme mass ratios \citep[for example, see][]{2017ApJS..230...15M}, we adopt a canonical flat mass ratio distribution. We stop the evolution of all binaries which initiate Roche-lobe overflow mass transfer. For our extreme mass ratios chosen at ZAMS it is expected that mass transfer would always become dynamically unstable, leading to a merger during the common envelope. Compared to the \POSYDON default\footref{fn:default} we use zero kicks for simplicity and the delayed prescription from \citep{2012ApJ...749...91F} to form BHs at all masses. In this prescription first a proto-neutron star is created and the fallback of envelope material onto it determines the mass of the remaining compact remnant \citep[see Section~8.3 in][for details]{2023ApJS..264...45F}. Finally, we allow for a wide range of initial orbital periods $\in [0.75,6\,000]\,\days$ to cover binaries with and without interaction at any of our initial masses.

\section{Results}
\label{sec:Results}

First, we highlight some relevant properties of our single star models calculated in \MESA, including those where we vary our default wind prescriptions, before presenting binary population synthesis results based on these single star models.

\subsection{Maximum Radius of Single, Massive Stars}
\label{sec:Radius}

\begin{figure}
  \centering
  \includegraphics[width=\columnwidth]{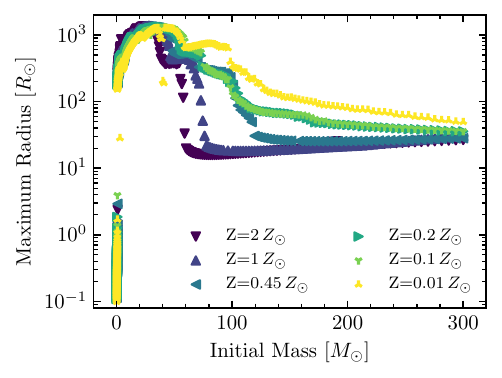}
  \caption{\label{fig:MaxRadius} The maximum radius of single stars as a function of their initial mass. The colour and symbol show different metallicities.}
\end{figure}

Naturally, massive stars have larger radii compared to their less massive counterparts on the ZAMS. While very low mass stars are often referred to as dwarfs, very massive stars are giants early on in their evolution. During their evolution stars usually expand, especially past the main sequence. Most of those expansion phases are a consequence of a mirroring effect that burning shells have due to their stabilising effect on the burning itself. The contraction of the core material leads to an expansion of the outer envelope layers above the burning region \citep[for a review see e.g.][]{2017use..book.....L}.

For very massive stars there is another effect which drives the expansion. The high luminosity produced by vigorous nuclear burning pushes a star's outer layers to larger radii. At the surface of these massive stars, this luminosity launches enhanced winds. These winds remove the outer stellar material, leading to a movement of the stellar photosphere to a lower radial coordinate. The combination of these two effects settle to an equilibrium radial coordinate for the stellar surface; in the case of very massive stars, especially at higher metallicity, the winds are so strong that material which would have led to an expansion of the stellar surface instead escapes from the star. This wind causes those stars to remain at a radius not far above the initial radius at ZAMS \citep[for a review see e.g.][]{1999isw..book.....L}.

Fig.~\ref{fig:MaxRadius} shows the maximal radial extension of a star of a given ZAMS mass during its evolution (limited to the age of the Universe $\sim 13.8\,\mathrm{G}\yr$, which causes a gap at very low masses, where stars would need more time to become giants) as a function of its initial mass. Depending on the mass at ZAMS there are different physical effects at work to determine the maximum radius. First, the more massive a star is the larger its radius. Second, the stars which evolve through shell burning will significantly expand during contraction phases of the core. These phases lead to the tip of the red giant branch. Subsequent burning phases can cause even larger radii on the asymptotic giant branch. This difference in the giant branches is the first and small step on the very left of Fig.~\ref{fig:MaxRadius} at a few $\Msun$ differentiating between a few hundred $\Rsun$ and several hundred or thousand $\Rsun$. This feature exists at all metallicities, while a larger abundance of heavier elements lead to a larger radius due to the higher resultant opacity. Depending on metallicity, for stellar masses in the range from $30$ to several hundreds of $\Msun$ the winds become so strong that the maximum radius is reached before the shell burning can lead to a maximum radius $>1\,000\,\Rsun$. The most massive stars reach their maximal extension already before they finish core-H burning. Transitions between wind prescriptions lead to several smaller features in Fig.~\ref{fig:MaxRadius}. For the \Gaia BHs to form without mass transfer the BH progenitor needs a maximum radius below a few hundred $\Rsun$.

\subsection{Final Mass of Single, Massive Stars}
\label{sec:Mass}

\begin{figure}
  \centering
  \includegraphics[width=\columnwidth]{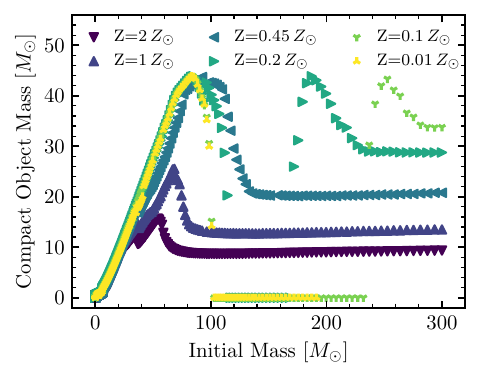}
  \caption{\label{fig:RemnantMass} The remnant mass function of single stars depending on their initial mass. The colour and symbol show different metallicities. We use the direct collapse prescription in \POSYDON. At \mbox{$Z\leq 0.2\,\Zsun$} stars with initial masses above $100\,\Msun$ might not form a BH because of pair instability and are shown here with a zero mass. It should be noted, that stars with initial masses above the pair instability gap form BHs outside the plotted range with masses above $100\,\Msun$ at the lowest shown metallicity.}
\end{figure}

Fig.~\ref{fig:RemnantMass} shows the final compact object mass using a simple direct collapse model to form the BH (alternative supernova prescriptions are shown in Appendix~\ref{sec:otherSN}). When comparing the ZAMS mass to the final BH mass, Fig.~\ref{fig:RemnantMass} exhibits a peak in the remnant mass which varies as a function of metallicity. Variations in the location of this peak are due to the amount of mass lost prior to collapse which varies with metallicity. At high metallicity the stellar winds are so strong that big parts, if not all, of the stellar envelope are lost during their evolution leading to relatively lower mass BHs \citep{bavera2023, 2024arXiv240707204V}. However, with their weaker winds, lower metallicity stars remain more massive and can enter the regime of pair instability, which ejects the envelope or even disrupts the whole star, leaving no remnant (zero mass in Fig.~\ref{fig:RemnantMass}). At metallicities below one fifth of the solar value, the early winds are too weak to strip massive stars and enable the WR winds. Here, strong winds connected to luminous blue variables have a similar effect, allowing the existence of a plateau in the BH mass above the pair instability regime, see Fig.~\ref{fig:RemnantMass}.

In the remainder of the paper we concentrate on the stellar winds. Winds produce a plateau in the BH mass over a wide range of high ZAMS masses. The typical BH mass at this plateau depends on the strength of the wind and therefore indirectly on metallicity as shown in Fig.~\ref{fig:RemnantMass}. Although the plateau is above $20\,\Msun$ for less than half the solar metallicity, plateau BHs would have typical masses below $10\,\Msun$ for twice the solar metallicity, for the default wind prescription in the \MESA models from \POSYDON{}~v2. Since \Gaia BH1 and BH2 are slightly sub-solar as most systems in the MW, throughout the rest of this work we focus only on solar-metallicity models.

\subsection{Variations of the Wind for Massive Stars}
\label{sec:Wind}

\begin{figure}
  \centering
  \includegraphics[width=\columnwidth]{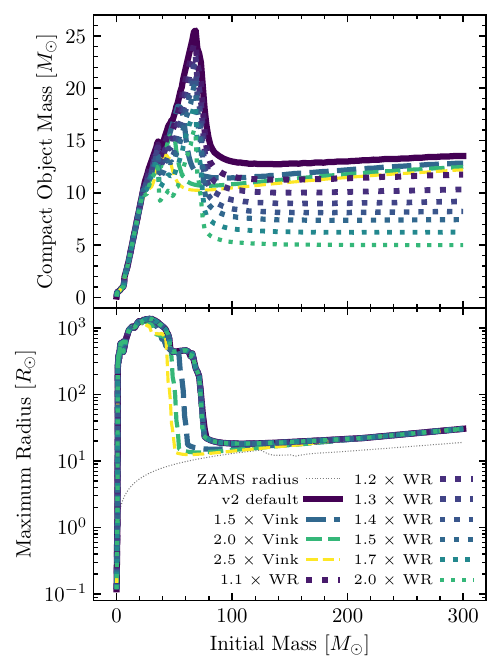}
  \caption{\label{fig:Wind} Top: like Fig.~\ref{fig:RemnantMass}; bottom: like Fig.~\ref{fig:MaxRadius}. But here we show different variations of the wind at solar metallicity, see legend. The grey dotted line showes the ZAMS radius, where the discontinuity at about $130\,\Msun$ is a numerical artefact in the ZAMS models due to the detection criterion of ZAMS. It should be noted, that the WR wind variations all overlap in the bottom panel.}
\end{figure}

During the evolution of a massive star at high metallicity there are two main wind regimes we want to investigate in more detail. Those are the Vink and the WR winds, see Section~\ref{sec:single} and Appendix~\ref{sec:wind_prescriptions} for more details.

The top panel of Fig.~\ref{fig:Wind} shows that the Vink wind strength determines the position of the maximum remnant mass peak. The stronger the Vink wind, the lower is the maximum mass and the lower is the ZAMS mass leading to the peak value. Beside the mass, the Vink wind determines when the star turns to become bluer and therefore define its maximum radius. Sufficiently strong winds remove the step at about $500\,\Rsun$, characterising the predominant difference between hot and cold winds in the bottom panel of Fig.~\ref{fig:Wind}. This step feature is also modified by other transitions of wind prescriptions, e.g. stronger winds when the star tries to enter the regime of luminous blue variables. An enhancement to the Vink winds have a minor effect on the most massive stars characterising the plateau region because those stars never evolve redder than the main sequence and therefore have a maximum radius very close to their ZAMS radius. A decrease of the Vink winds \citep[like suggested by][]{2014ARA&A..52..487S, 2022A&A...665A.133G} would have similar effects as going towards lower metallicity, see Sections~\ref{sec:Radius} and \ref{sec:Mass}.

The WR wind does not effect the ZAMS mass leading to the peak in BH mass, but it can somewhat lower the resulting BH mass at the peak. All stars that keep their H-rich envelope throughout their evolution, never experience this kind of wind. Additionally, the WR wind is responsible for the formation of the plateau due to its luminosity scaling. Therefore, the typical BH mass in the plateau regime strongly depends on the WR wind prescription adopted. We point out that usually, the maximum radius of a star is reached before this wind kicks in, thus it has no effect on whether binary interactions will be triggered or not (see the bottom panel of Fig.~\ref{fig:Wind}).

\begin{figure}
  \centering
  \includegraphics[width=\columnwidth]{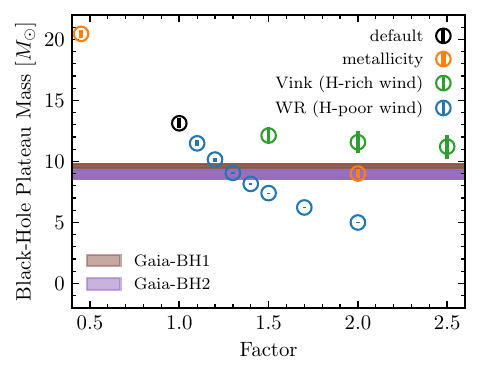}
  \caption{\label{fig:BHplateau} The range of BH masses in the plateau regime depending only on the metallicity in $\Zsun$ (orange, $0.45$ and $2\,\Zsun$), Vink-wind factor (green), or WR-wind factor (blue), while the others are at default. The default at solar metallicity and without any wind modification is shown in black.}
\end{figure}

In Fig.~\ref{fig:BHplateau} we show how the plateau mass varies with different wind prescriptions and metallicities. Here, we define the plateau as the mass range between the minimum compact-object mass at higher ZAMS masses than the peak, and the maximum at even higher masses than the aforementioned minimum, which usually coincides with the \MESA model with the largest ZAMS mass. The plateau mass depends only weakly on the Vink wind applied during the main sequence of massive stars. The optically thick WR wind applied after the H-rich envelope is lost has a clear impact on the typical BH mass in the plateau regime, see Fig.~\ref{fig:BHplateau}. Additionally, its strength causes the plateau mass range to become narrower. To have the plateau in the mass range of the \Gaia BHs our simulations of solar metalicity would require a boost of the WR winds by a factor in the range $[1.15, 1.35]$ depending on the supernova prescription. Some trends are slightly modified when using different prescriptions for the BH formation, cf. Appendix~\ref{sec:otherSN}.

\subsection{Population Models of Non-interacting Binaries}
\label{sec:Pop-syn}

\begin{figure*}
  \centering
  \includegraphics[width=\textwidth]{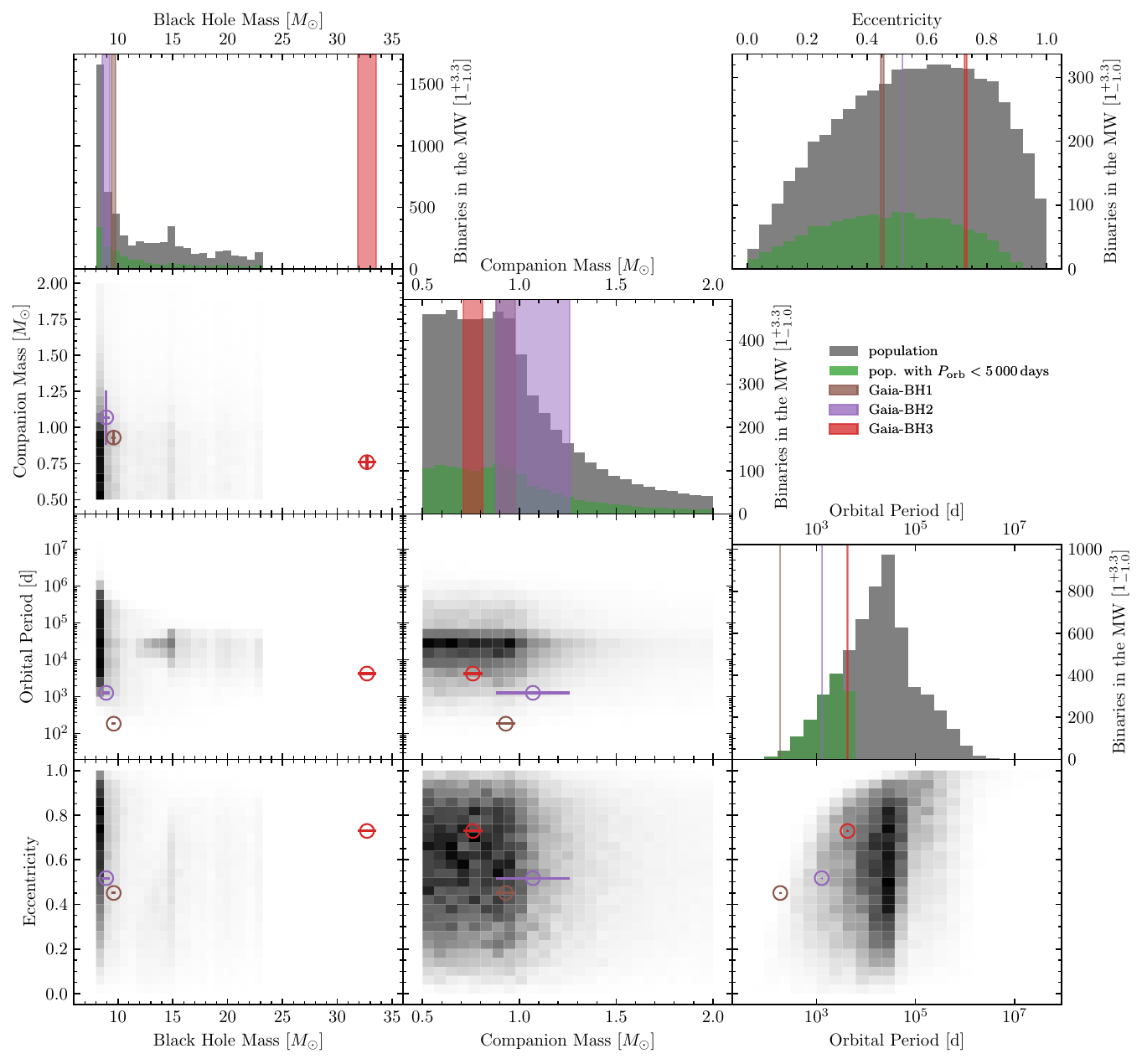}
  \caption{\label{fig:Pop-syn} A corner plot showing the black hole mass, the companion mass, the orbital period and the eccentricity of binaries in a population synthesis run with enhanced WR wind (factor $=1.2$). While the grey histograms are the full population, the green one are limited to binaries with orbital periods up to $5\,000\,\days$. The observationally determined values for the \Gaia black holes are shown in colour \citep{el-badry-BH1, el-badry-BH2, Gaia-BH3}. It should be noted, that the histograms are weighted to represent the number of such binaries in the MW, which has large uncertainties as given by the unity unit $[1^{+3.3}_{-1.0}]$ -- for details see Appendix~\ref{sec:norm}.}
\end{figure*}

Using our adapted version of \POSYDON, we have performed several population synthesis simulations to inspect the distribution of detached BH binary properties with a low-mass companion in wide-period orbits, such as the \Gaia BH systems. Fig.~\ref{fig:Pop-syn} shows an example intrinsic population, without accounting for any \Gaia selection effects, using the setup outlined in Section~\ref{sec:Methods}. The observed \Gaia BHs are over-plotted in colour. It should be noted that we do not try to reproduce \Gaia BH3 here, because it clearly formed at lower metallicity, which is required to form BHs of this mass without binary interactions. Nevertheless, we have included \Gaia BH3 in Fig.~\ref{fig:Pop-syn} to show that systems with similar orbital parameters and only lower BH mass are formed. Hence, it is expected that \Gaia BH3 could potentially form at lower metallicity without binary interactions as well. On the other hand, very low mass BHs, neutron stars or compact objects in the mass gap \citep[like the one recently proposed by][]{2024arXiv240906352S} would require very high metallicity or extremely strong winds to form without binary interactions.

A typical progenitor of \Gaia BH2 starts in our simulation with stars of $92.07$ and $1.092\,\Msun$ in a $15.13\,\days$ orbit and an eccentricity of $0.4796$ at ZAMS. Prior to collapse the primary loses most of its mass in winds reducing its mass down to $11.12\,\Msun$, which will form a $9.446\,\Msun$ BH. The mass loss due to winds and the BH formation cause the system to widen to $878.9\,\days$ prior and $1\,275\,\days$ post collapse. While the eccentricity stays nearly unchanged during the progenitor evolution it becomes $0.5168$ after the BH formation. While the companion evolves, the system is observable for $>8\,\mathrm{G}\yr$, where tides start to circularize the binary after the companion left its main sequence. Finally the system becomes a wide low mass X-ray binary.

For the population, there is a peak caused in the BH mass function which originated from the plateau region (see BH masses $\lesssim 10\,\Msun$ in the top left panel of Figs~\ref{fig:Pop-syn} and \ref{fig:Pop-syn2}). Depending on the strength assumed for kicks at BH formation there can be a second, usually smaller, peak at the BH mass which corresponds to the lowest massive stars (at initial mass of $\lesssim 50\,\Msun$ leading to a $\approx 15\,\Msun$ BH), which reach a maximum radius of about $500\,\Rsun$. Hence, this peak requires a larger orbital period than the one of the plateau. In the two dimensional heat map (first column of the third row in Figs~\ref{fig:Pop-syn} and \ref{fig:Pop-syn2}) between BH mass and orbital period one can easily differentiate between the two sub populations. It should be noted, that none of the \Gaia BHs could belong to the second sub population, which requires a period of about $20\,000\,\days$.

The companion mass shows a clear preference toward lower masses. This preference is simply caused by their longer life time making lower-mass systems statistically more likely. Slightly below $1\,\Msun$ this trend levels off because the stellar lifetime reaches the age of the Universe which is the maximum allowed in our population synthesis. In observations, companions slightly below $1\,\Msun$ are most likely to be found first because of being the brightest under the most common companions. Companions with masses of $2\,\Msun$ are more than an order of magnitude less common. The power law behaviour suppresses companions with larger masses even more, thus resulting in a very small sample that will be very rare to observe in the MW.

The only other two-dimensional representation that shows a correlation is the one between the orbital period and the eccentricity. The structure is a result of the mass loss at BH formation. The reported BHs observed by \Gaia are at the lower end of the period distribution.

\subsection{Detectability by \Gaia}
\label{sec:GaiaSelectionFunction}

\begin{figure}
  \centering
  \includegraphics[width=\columnwidth]{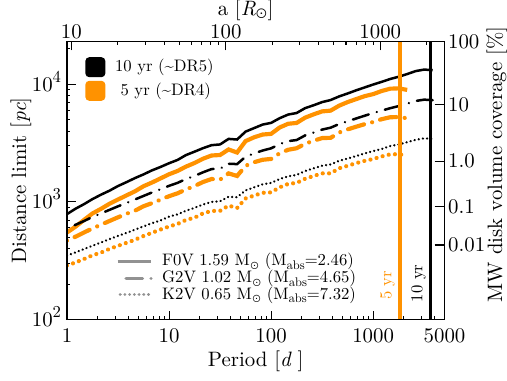}
  \caption{\label{fig:GaiaSensibility} \Gaia detection limits for a non-luminous companions of $10\,\Msun$ on a circular orbit around a F0V, G2V and K2V star. The lines represent the (sky-uniform) median distance limit at which the $\Delta \chi^2=100$ between a single star and circular Keplerian orbital model fit, corresponding to (typically) well-constrained orbital fits. Estimates provided for the $5\,\yr$ nominal mission ($\sim$DR4) and $10\,\yr$ extended mission ($\sim$DR5) as function of period. Successful detections drop quickly for orbital periods longer than the data time-span, hence the graphs are truncated at $5$ and $10\,\yr$. On top, the semi-major axis ($a$) is specified for a G2V star, though the difference for the other masses would be hardly noticeable. On the right side we provide a rough estimate of the Milky Way coverage for a specific distance limit, assuming a disk of height $1.15\,\kpc$ and radius $15\,\kpc$, hence we get a full coverage at $23\,\kpc$. All detection values are assuming zero extinction.}
\end{figure}

Although a detailed modelling and analyses of the observability of our sample by \Gaia is beyond the scope of this paper, it is instructive to examine the \Gaia sensitivity function in the visible star's mass-range we simulated. The vast majority of orbital companions with orbital periods $>10\,\days$ will be identified in the astrometric data of \Gaia \citep[see e.g.][]{2024NewAR..9801694E}. We illustrate the expected \Gaia astrometric sensitivity function in Fig.~\ref{fig:GaiaSensibility} for $5\,\yr$ ($\sim$DR4) and $10\,\yr$ ($\sim$DR5), showing the sky-averaged detectability horizon for unseen objects on circular orbit around companion stars of mass $0.65$, $1.02$, and $1.59\,\Msun$ \citep[following appendix~A of][]{2022A&A...661A.151H}. As expected we see a higher sensitivity (i.e., more distant detection horizon) for longer periods, though limited to roughly the time-span of the observations being roughly $1\,000$, $2\,000$, and $4\,000\,\mathrm{d}$ for DR3, DR4, and DR5, respectively. Comparing this to the orbital period distribution of Fig.~\ref{fig:Pop-syn}, we see that with each succeeding data release \Gaia will cover more and more of the expected orbital period distribution. Thus, each release is expected to increase the number of \Gaia BHs. Additionally, each successive data release will double the number of data points, increasing the overall signal to noise ratio as can be seen by the offset between the $5$ and $10\,\yr$ lines of Fig.~\ref{fig:GaiaSensibility}. 

We note that the eccentricity distribution of the systems shown in Fig.~\ref{fig:Pop-syn} is far from zero and has a typical value of $0.4-0.8$ like the first \Gaia BHs. Illustrated in figure~A.1 of \citet{2022A&A...661A.151H}: higher eccentricity reduces the detectability slowly: with a distance limit drop of about $10\%$ at $e=0.5$ and only exceeding a drop of more than $20\%$ beyond $e=0.7$. 
This and Fig.~\ref{fig:GaiaSensibility} describe the all-sky \textit{median} behaviour, while in reality the detectability as function of all astrometric orbital parameters can differ strongly as function of sky position due to the \Gaia scanning law.

\section{Discussion and Conclusions}
\label{sec:Discussion}

In this paper, we show that BHs found in wide binaries, like \Gaia BH1 and \Gaia BH2, can form via a non-interacting, isolate binary evolution channel at higher, solar-like metallicities. The clearest signature of the presented formation channel is a preferred BH mass in wide binaries with low-mass companions at a given metallicity. Additionally, if the eccentricity distribution of systems like the \Gaia BHs is similar to samples of young systems a formation channel with small interaction (e.g. tides) and low kicks might be the favoured one. The verification that such a channel is dominant would require a significantly larger observational sample.

If it can be demonstrated that the \Gaia BHs formed without binary interaction, the masses of the BH and the metallicity of the companion can be used to constrain the winds of massive stars, especially the WR wind. We find a best match of \Gaia BHs for a boosting factor of about $1.2$ on the WR winds. Nevertheless, there are uncertainties not covered by our simple models, e.g. a different scaling of the winds, or the impact of the BH formation prescription -- the latter is less important for BHs plateaus above $10\,\Msun$, see Appendix~\ref{sec:otherSN}. Such degeneracies can only be overcome with a large sample of wide binaries containing a BH and a low mass companion. Finally, it should be noted, that we only measure where the final mass settles. Hence, we cannot differentiate between stronger mass loss during a shorter period of time versus moderate mass loss over a longer time period, e.g. different prescriptions for BH formation favour different wind strength to accumulate at the same mass range.

Our simulations predict a second, smaller peak in the BH mass distribution of wide, non-interacting BH binaries (cf. $\approx 15\,\Msun$ in Fig.~\ref{fig:Pop-syn}). This local overabundance should coincide with periods always above $1\,000\,\days$. There, the progenitors are stars with ZAMS masses below the mass of the ones forming the most massive BHs at a given metallicity. Such stars experienced at most a very short phase with WR winds in their evolution. Thus, the final BH mass from this subpopulation responsible for the second peak probes the early winds and radial expansions of stars. However, the small contribution of this secondary sub-populations and the long periods of its binaries make it unlikely that it can be directly contained from Gaia observations.

Other modifications like enhance overshooting \citep{2024MNRAS.535L..44G} have very similar effects like our changes on the Vink wind. Changes on the overshooting or other mixing processes can enhance the luminosity on the main sequence and therefore cause larger winds as well. As shown in Fig.~\ref{fig:BHplateau} and \citet{2024MNRAS.535L..44G} rather extreme changes are required to get BHs in the reported mass range of the first two \Gaia BHs. Such a strong push of the main sequence towards higher luminosities is not observed to date.

It is very difficult to compare this channel to other binary formation channels, which probably coincide in nature, because of the extreme mass ratio it requires already early on the ZAMS. In this regime observational samples are notoriously incomplete precluding reliable estimates of the prevalence of initial binaries at ZAMS with the extreme mass ratios required to form \Gaia BH-like binaries. Nevertheless, adopting a flat mass ratio distribution, which is a pure extrapolation, we estimate that hundreds of systems like \Gaia BH1 and BH2 should exist in the MW, see bottom rows of Table~\ref{tab:MWsys}. Furthermore, our models predict many more systems at larger orbital periods which are difficult to find given \Gaia's limited lifetime. Nevertheless, we expect many more systems to be found in future \Gaia data releases covering a longer time baseline.

\begin{acknowledgements}
This work was supported by the Swiss National Science Foundation (project number PP00P2\_211006 ).
The POSYDON project is supported primarily by two sources: the Swiss National Science Foundation (PI Fragos, project numbers PP00P2\_211006 and CRSII5\_213497) and the Gordon and Betty Moore Foundation (PI Kalogera, grant award GBMF8477). 
J.J.A.~acknowledges support for Program number (JWST-AR-04369.001-A) provided through a grant from the STScI under NASA contract NAS5-03127. 
M.B. ackowledges support from the Boninchi Foundation.
K.K. acknowledges support from the Spanish State Research Agency, through the María de Maeztu Program for Centers and Units of Excellence in R\&D, No. CEX2020-001058-M. 
K.A.R.\ is also supported by the Riedel Family Fellowship and thanks the LSSTC Data Science Fellowship Program, which is funded by LSSTC, NSF Cybertraining Grant No.\ 1829740, the Brinson Foundation, and the Moore Foundation; their participation in the program has benefited this work.
Z.X. acknowledges support from the China Scholarship Council (CSC).
E.Z. acknowledges funding support from the Hellenic Foundation for Research and Innovation (H.F.R.I.) under the "3rd Call for H.F.R.I. Research Projects to support Post-Doctoral Researchers" (Project No: 7933). 
\end{acknowledgements}

\bibliography{refs}{}
\bibliographystyle{aa}

\begin{appendix} 

\section{Wind Prescriptions}
\label{sec:wind_prescriptions}

Here, we are summarising the wind prescriptions used in \POSYDON's default adoptions to winds in \MESA.

We calculate a ``cool'' wind for low photospheric temperatures, \mbox{$T_\mathrm{photo}\leq 12\,000\,\kelvin$}. If the central He abundance is smaller than $10^{-6}$ and the difference between He and C core is small, \mbox{$M_\mathrm{He, core}-M_\mathrm{C, core}<0.1\,\Msun$}, hence for cool asymptotic giant branch stars, we use the ``Blocker'' wind scheme. Otherwise, for cool red giant branch stars, we use the ``Dutch'' wind scheme.
For \mbox{$T_\mathrm{photo}\geq 8\,000\,\kelvin$} we calculate a ``hot'' wind using the ``Dutch'' wind scheme (initial mass \mbox{$M_\mathrm{ini}>8\,\Msun$}) or ``Reimers'' wind (\mbox{$M_\mathrm{ini}\leq 8\,\Msun$}).
For \mbox{$T_\mathrm{photo}<8\,000\,\kelvin$} we use the cold wind only and for \mbox{$T_\mathrm{photo}>12\,000\,\kelvin$} we use the hot wind only. In the intermediate range we interpolate between the two winds linearly with $T_\mathrm{photo}$.

In the ``Blocker'' wind scheme, we use the maximum of $0.2$ times the value of equation~(2) in \citet{1995A&A...299..755B} and our ``Reimers'' wind -- including the scaling factor.
In the ``Dutch'' wind scheme \citep{2009A&A...497..255G}, for \mbox{$T_\mathrm{photo}\leq 10\,000\,\kelvin$} we use ``de~ Jager'' wind, for \mbox{$T_\mathrm{photo}\geq 11\,000\,\kelvin$} we apply ``Nugis \& Lamers'' wind if the surface H abundance, $X<0.4$ and ``Vink'' wind scheme for larger H abundances, otherwise we linearly interpolate with $T_\mathrm{photo}$.

In the ``Reimers'' wind prescription, we adopt a wind mass loss rate according to \citet{1975psae.book..229R} with an additional scaling factor of $0.1$.
In the ``de~Jager'' wind prescription, we use the linear approximation of the empirical interpolation given in equation~(1) of \citet{1988A&AS...72..259D}.
In the ``Vink'' wind prescription, we use equations (24) and (25) of \citet{2001A&A...369..574V} with equations (14) and (15) to make the transition between the high- and low-temperature regime. Throughout this work, we modify the Vink wind prescription by including a multiplicative factor to the final wind mass loss rate, see Section~\ref{sec:Wind}. In the ``Nugis \& Lamers'' wind prescription, we use equation~(1) of \citet{2000A&A...360..227N}, where we use the surface values of the He abundance and \mbox{$Z=1-X-Y$}. In our wind variations we refer to this wind as Wolf-Rayet (WR) wind and use a simple multiplicative factor in front of the usual wind value, see Section~\ref{sec:Wind}.

If a star crosses the Humphreys-Davidson limit \citep[\mbox{$L>6\times 10^{5}\,\Lsun$} and 
\mbox{$R\,\Rsun^{-1}>10^5\,L^{-0.5}\,\Lsun^{0.5}$}, 
][]{1979ApJ...232..409H, 1994PASP..106.1025H}, is not on the thermally pulsating asymptotic giant branch, and has a surface H abundance $X>0.1$, then we apply luminous blue variable winds which is taken to be the maximum of $10^{-4}\,\Msun\yr^{-1}$ and the otherwise calculated wind.

\section{Other Prescriptions for the Compact Object formation}
\label{sec:otherSN}
\begin{figure}
  \centering
  \includegraphics[width=\columnwidth, clip, trim=0px 33px 0px 0px]{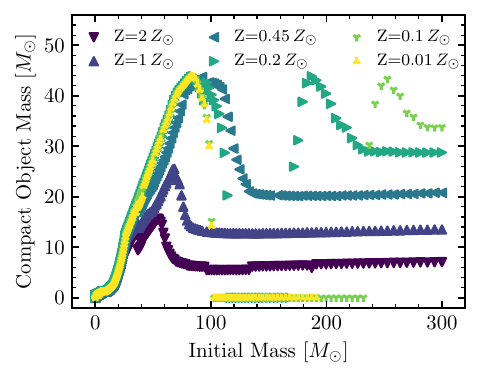}
  \includegraphics[width=\columnwidth, clip, trim=0px 0px 0px 7px]{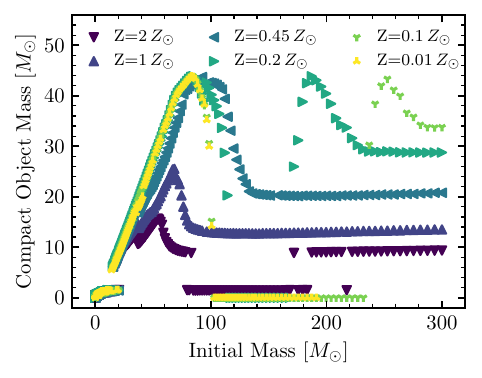}
  \caption{\label{fig:RemnantMass3} Like Fig.~\ref{fig:RemnantMass}, but showing top: Fryer+12-delayed \citep{2012ApJ...749...91F}; bottom: Patton+Sukhbold20-engine \citep{2020MNRAS.499.2803P} prescription for the compact object formation.
  }
\end{figure}

\begin{figure}
  \centering
  \includegraphics[width=\columnwidth, clip, trim=0px 33px 0px 0px]{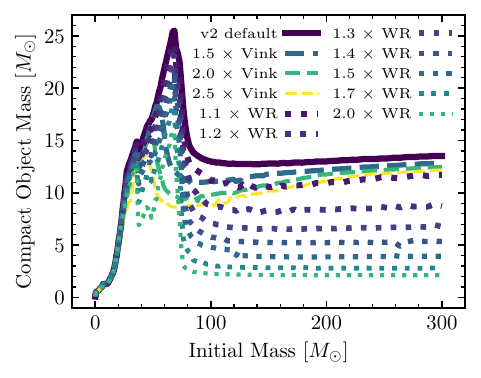}
  \includegraphics[width=\columnwidth, clip, trim=0px 0px 0px 7px]{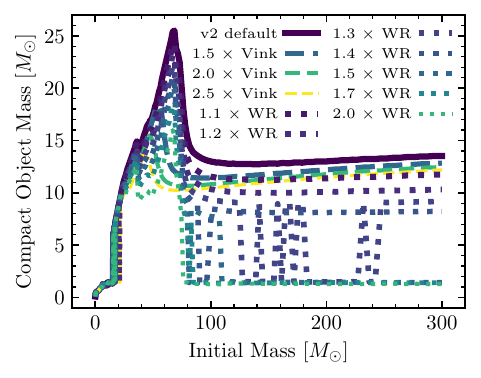}
  \caption{\label{fig:RemnantMass4} Like the top panel of Fig.~\ref{fig:Wind}, but showing top: Fryer+12-delayed \citep{2012ApJ...749...91F}; bottom: Patton+Sukhbold20-engine \citep{2020MNRAS.499.2803P} prescription for the compact object formation.
  }
\end{figure}

\begin{figure}
  \centering
  \includegraphics[width=\columnwidth, clip, trim=0px 32px 0px 0px]{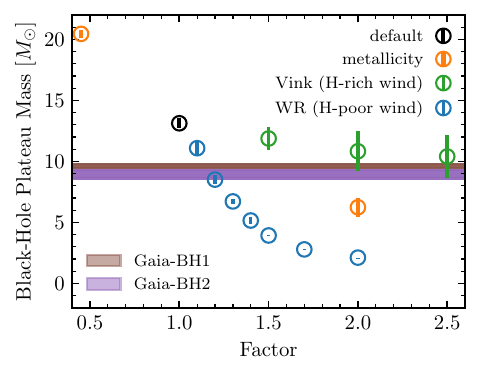}
  \includegraphics[width=\columnwidth, clip, trim=0px 0px 0px 7px]{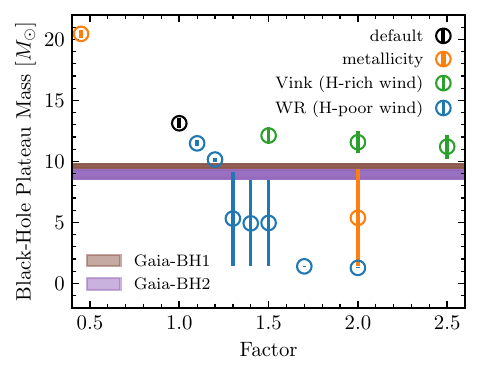}
  \caption{\label{fig:BHplateau5} Like Fig.~\ref{fig:BHplateau}, but showing top: Fryer+12-delayed \citep{2012ApJ...749...91F}; bottom: Patton+Sukhbold20-engine \citep{2020MNRAS.499.2803P} prescription for the compact object formation. It should be noted, that here the plateau spans over the mass gap between neutron stars and BHs for the strongest wind variations.}
\end{figure}

In Figs~\ref{fig:RemnantMass3} to \ref{fig:BHplateau5} we show the impact of alternative prescriptions for the BH formation. Following the delayed prescription of \citet{2012ApJ...749...91F} looks very similar to the simple direct collapse model (Fig.~\ref{fig:RemnantMass} and top panel of Fig.~\ref{fig:Wind}), where only the lower compact object masses are shifted due to differences in the amount of fallback. On the other hand the prescription based on explodability \citep{2020MNRAS.499.2803P} may form neutron stars instead of BHs for some models in the plateau range. It should be noted that it is unclear whether this prescription is valid for very massive stars which got highly stripped.

\subsection{The Maximum Black Hole Mass}
\begin{figure}
  \centering
  \includegraphics[width=\columnwidth]{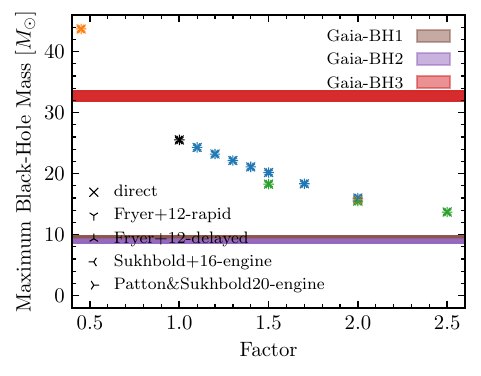}
  \caption{\label{fig:BHpeak} The maximum BH mass depending on the metallicity in $\Zsun$ (orange), Vink-wind factor (green), or WR-wind factor (blue). The default at solar metallicity and without any wind modification is shown in black. The different symbols indicate the prescription for the BH formation.}
\end{figure}

Fig.~\ref{fig:BHpeak} shows how the maximum BH mass scales with the metallicity and additional factors on the Vink or WR wind. This is in agreement with the findings of wind variations on the maximum mass of a BH formed at solar metallicity by \citet{bavera2023}. Because the peak masses are always well above $10\,\Msun$ the different prescriptions of BH formation in \POSYDON lead to the same results.

\subsection{The Black Hole Progenitor}
\begin{figure}
  \centering
  \includegraphics[width=\columnwidth]{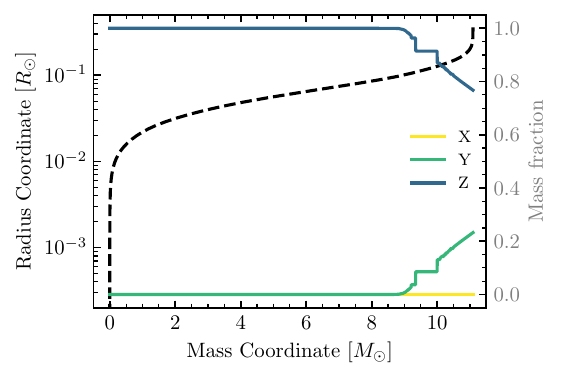}
  \caption{\label{fig:BHprogenitor} Final profile of an initially $92.42\,\Msun$ star with WR winds enhanced by a factor $1.2$. In black, dashed the radius and in color, solid the mass fractions of hydrogen (X), helium (Y) and metals (Z).}
\end{figure}

In Fig.~\ref{fig:BHprogenitor} we show the final profile at core carbon depletion similar to the example star in section~\ref{sec:Pop-syn} as a typical case from the BH mass plateau. This profiles is used to determine the BH after collapse. There is no hydrogen left in the star and the winds have even stripped off most of the helium rich material. The final BH consists of all the carbon-oxygen core and a small amount of helium will fall back into the final remnant of about $9.4\,\Msun$.

\section{Another Population run}
\label{sec:Pop-syn2}
\begin{figure*}
  \centering
  \includegraphics[width=\textwidth]{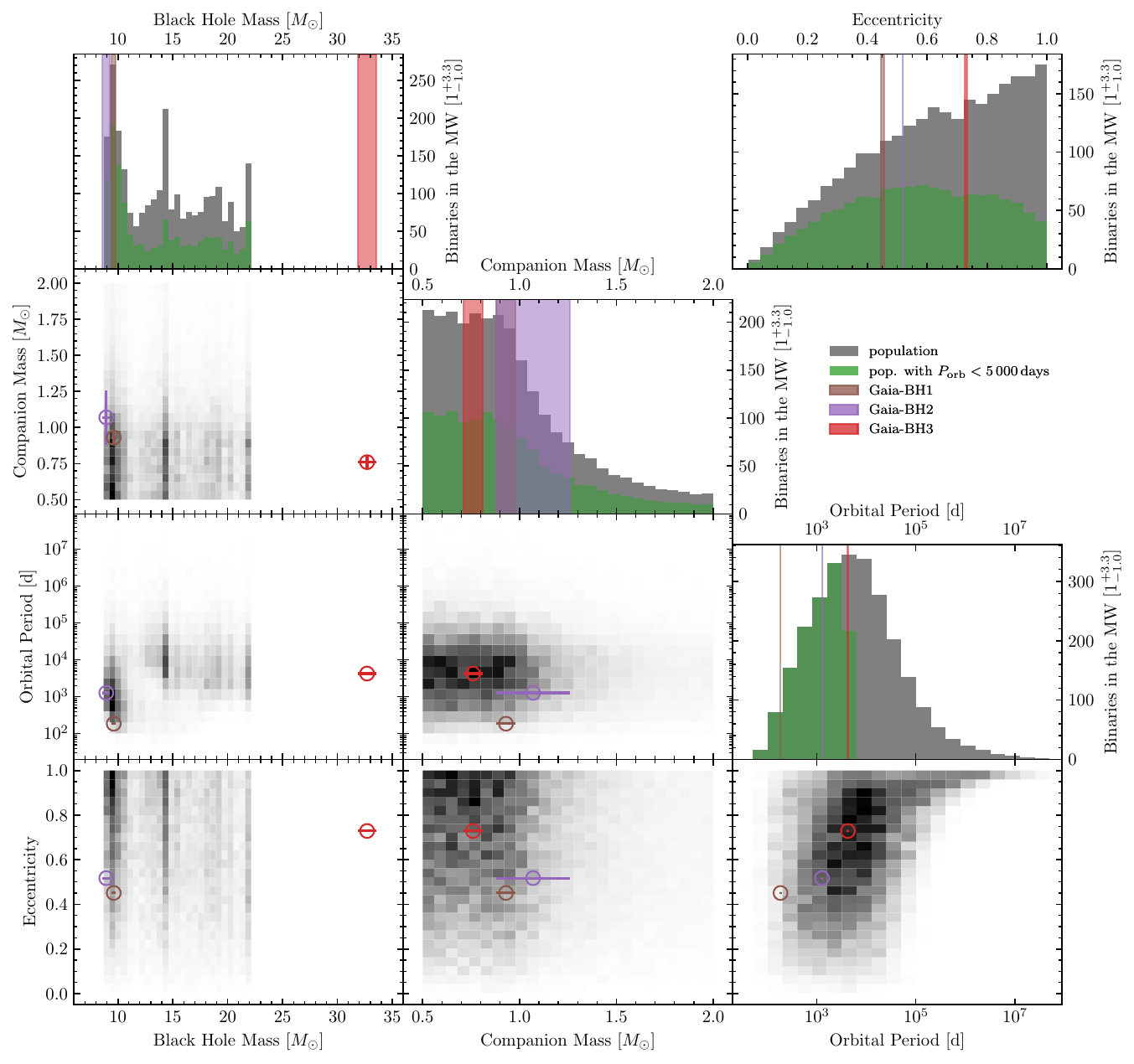}
  \caption{\label{fig:Pop-syn2} A corner plot showing the black hole mass, the companion mass, the orbital period and the eccentricity of binaries in a default \POSYDON population synthesis run with enhanced WR wind (factor $=1.3$), see text for more details. While the grey histograms are the full population, the green one are limited to binaries with orbital periods up to $5\,000\,\days$. The observationally determined values for the \Gaia black holes are shown in colour \citep{el-badry-BH1, el-badry-BH2, Gaia-BH3}. It should be noted, that the histograms are weighted to represent the number of such binaries in the MW, which has large uncertainties as given by the unity unit $[1^{+3.3}_{-1.0}]$ -- for details see Appendix~\ref{sec:norm}.}
\end{figure*}

Fig.~\ref{fig:Pop-syn2} show the results from a population synthesis run with several changes compared to Fig.~\ref{fig:Pop-syn}. Those are: 1) using stronger WR winds (factor$=1.3$), 2) using Patton+Sukhbold20-engine \citep{2020MNRAS.499.2803P} for BH formation, 3) using kicks scaled by one over the BH mass, 4) using zero eccentricity distribution at ZAMS.

With a different engine to calculate the final BH mass, a different wind strength is needed to reproduce the BH masses of \Gaia BH1 and BH2. Still, \Gaia BH3 is above the maximum mass of BHs formed at solar metallicity. Using kicks at BH formation leads to significant orbital changes when the BH is born. Hence, the choice of initial eccentricity distribution is less important and higher eccentricities are more common. Additionally, the kick leaves an imprint in the period-eccentricity plane. The other key findings of Section~\ref{sec:Pop-syn} hold here as well.

\section{Normalising the Populations}
\label{sec:norm}
\begin{table*}
    \caption{\label{tab:MWsys}Counts of systems in the MW for different selections on the BH mass ($M_1$), the companion mass ($M_2$), the orbital period ($P_\mathrm{orb}$), and the eccentricity ($e$). The two simulations (sim1 and sim2) are the ones shown in Figs~\ref{fig:Pop-syn} and \ref{fig:Pop-syn2}, respectively.}
    \centering
    \begin{tabular}{c|c|c|c|c|c}
        $M_1 [\Msun]$ & $M_2 [\Msun]$ & $P_\mathrm{orb} [\mathrm{d}]$ & e & sim1 & sim2 \\
        \hline
        all & all & all & all & $5805^{+19156}_{-5805}$ & $2619^{+8643}_{-2619}$ \\
        $[9.5, 10.5]$ & $[0.50, 2.00]$ & $[100.00, \phantom{0}300.0]$ & all & $\phantom{000}9^{+31\phantom{000}}_{-9}$ & $\phantom{00}50^{+166\phantom{0}}_{-50}$ \\
        $[8.6, \phantom{0}9.8]$ & $[0.88, 1.26]$ & $[185.54, 1277.3]$ & $[0.446, 0.5185]$ & $\phantom{000}6^{+19\phantom{000}}_{-6}$ & $\phantom{000}4^{+13\phantom{000}}_{-4}$ \\
        all & all & $<5\,000$ & all & $1391^{+4592\phantom{0}}_{-1391}$ & $1293^{+4267}_{-1293}$ \\
        $[8.6, \phantom{0}9.8]$ & $[0.88, 1.26]$ & all & all & $\phantom{0}429^{+1414\phantom{0}}_{-429}$ & $\phantom{0}146^{+482\phantom{0}}_{-146}$ \\
        $[8.6, \phantom{0}9.8]$ & $[0.88, 1.26]$ & $<5\,000$ & all & $\phantom{0}123^{+406\phantom{00}}_{-123}$ & $\phantom{0}100^{+331\phantom{0}}_{-100}$ \\
    \end{tabular}
\end{table*}
To get numbers of binaries in the MW from our simulations, we have to normalise our counts.
\begin{equation}
  N_\mathrm{MW} = f_\mathrm{bin} \, f_\mathrm{mass} \, f_\mathrm{MW} \, N_\mathrm{sim},
\end{equation}
where \mbox{$f_\mathrm{bin} = 0.7$} is the assumed initial overall binary fraction\footnote{\label{fn:v1_default}Default values taken from \POSYDON v1 \citep{2023ApJS..264...45F}.}. For the initial primary mass distribution we follow \citet{2001MNRAS.322..231K}, \mbox{$\xi(m) \propto m^{-\alpha_i}$} with
\begin{equation}
 \begin{split}
  \alpha_0 = 0.3 \pm 0.7 \quad&\text{for}\quad 0.01\,\Msun \leq m < 0.08\,\Msun\\
  \alpha_1 = 1.3 \pm 0.5 \quad&\text{for}\quad 0.08\,\Msun \leq m < 0.50\,\Msun\\
  \alpha_2 = 2.3 \pm 0.7 \quad&\text{for}\quad 0.50\,\Msun \leq m < 1.00\,\Msun\\
  \alpha_3 = 2.3 \pm 0.7 \quad&\text{for}\quad 1.00\,\Msun \leq m.
 \end{split}
\end{equation}
The secondary mass we infer from a mass ratio distribution, \mbox{$p_q \propto q^{\gamma}$}. This we assume to be flat, hence \mbox{$\gamma = 0$} for all mass ratios. Thus we get a correction factor\footref{fn:v1_default}
\begin{equation}
  f_\mathrm{mass} = \frac{\int_{20}^{150}\int_{0.5/m}^{2.0/m}\xi(m)\,p_q(q)\,\mathrm{d}q\,\mathrm{d}m}{\int_{0.01}^{200}\int_{0}^{1}\xi(m)\,p_q(q)\,\mathrm{d}q\,\mathrm{d}m}.
\end{equation}
We obtain our uncertainty for the dominating factor by having different assumptions for the distribution of \mbox{$q < 0.1$}. First, there are no binaries with low mass ratio, \mbox{$p_q = 0 \Rightarrow N_\mathrm{MW,min}=0$}. Second, all single stars are actually binaries with low mass ratio, \mbox{$\int_0^{0.1} p_q\,\mathrm{d}q = 0.3$} and \mbox{$f_\mathrm{bin} = 1$}. Thus we have \mbox{$N_\mathrm{MW,max}\approx 4.3 \, N_\mathrm{MW}$}.

We normalise to a MW like galaxy by comparing the simulated mass to the stellar mass of the MW, \mbox{$M_\mathrm{MW} = 5\times 10^{10}\,\Msun$} \citep[][where we optimistically assume that all binaries have formed at solar metallicity]{2007ApJ...662..322H, 2009A&A...505..497Y}.
\begin{equation}
  f_\mathrm{MW} = \frac{M_\mathrm{MW}}{M_\mathrm{sim}}
\end{equation}
Finally, we correct the simulated number of binaries already to account for the lifetime of systems like the \Gaia BHs.
\begin{equation}
  N_\mathrm{sim} = \sum_\mathrm{sim} \frac{t_\mathrm{selected}}{t_\mathrm{max}}
\end{equation}
We select \Gaia BH like systems as binaries containing a BH with a detached companion. This provides us with the time those binaries spend in this phase, $t_\mathrm{selected}$. Each binary is in maximum evolved for the age of the Universe\footref{fn:v1_default}, \mbox{$t_\mathrm{max}=13.8\,\mathrm{Gyr}$}, hence \mbox{$t_\mathrm{selected} < t_\mathrm{max}$}.

Table~\ref{tab:MWsys} summarises counts of systems in the MW for different selections on our simulates samples. The first one includes all simulated binaries resulting in a detached binary with a BH and a low mass companion, which did not interact during the evolution. The second line tries to mimic the selection done in \citet{el-badry-BH1}. The third selection unifies the parameter ranges to just include \Gaia BH1 and BH2 at the same time. The fourth line represents the green histograms in Figs~\ref{fig:Pop-syn} and \ref{fig:Pop-syn2}. The last two join the mass estimates of \Gaia BH1 and BH2, while being more generous on the orbital parameters.

\section{Milky Way Disk Coverage}
\label{sec:conv}
For the volume coverage, we used a simple geometric calculation to convert the distance to a coverage of the MW disk with a thickness of $2.3\,\kpc$ and a radius of $15\,\kpc$ \citep{2007ApJ...662..322H, 2009A&A...505..497Y}. It should be noted that \Gaia is not in the centre of the disk and we used a solar system distance of about $8\,\kpc$ to the disk centre. Thus, we can use a spherical volume observed by \Gaia for distances up to half the thickness. At larger distances we had to remove the two spherical sections above and below the disk, which hold up to about $7\,\kpc$. Finally, we know that we would have the full disk covered within about $23\,\kpc$. All this still assumes no extinction, which does not hold towards the galactic centre. To combine the population synthesis results (normalised by mass) with the volume coverage, one would need to assume a constant average density.

\end{appendix}

\end{document}